\documentclass[aps,prd,twocolumn,superscriptaddress,nofootinbib,floatfix,longbibliography]{revtex4-2}
\pdfoutput=1

\usepackage{amsmath}
\usepackage{amsfonts}
\usepackage{amssymb}
\usepackage{mathrsfs}
\usepackage{graphicx}
\usepackage{xcolor}
\usepackage{hyperref}
\usepackage{multirow}
\usepackage{slashed}
\usepackage{epsf}
\usepackage{slashed} 

\graphicspath{{new_Plots/}}

\newcommand{\mgrav}{m_{3/2}}
\newcommand{\mzero}{m_0}

\newcommand{\mchi}{m_{\tilde\chi_1^0}}
\newcommand{\mchar}{m_{\tilde\chi_1^\pm}}

\newcommand{\tanb}{\tan\beta}
\newcommand{\gev}{\, {\rm GeV}}
\newcommand{\tev}{\, {\rm TeV}}

\begin{document}

\title{
Probing the Supersymmetric Grand Unified Theories with Anomaly Mediation at the Future Proton-Proton Colliders and Hyper-Kamiokande Experiment
}

\author{Waqas Ahmed}
\email{waqasmit@hbpu.edu.cn}
\affiliation{School of Mathematics and Physics, Hubei Polytechnic University, Huangshi 435003, China }
\author{Tianjun Li}
\email{tli@mail.itp.ac.cn}
\affiliation{School of Physics, Henan Normal University, Xinxiang 453007, P. R. China}
\author{Shabbar Raza}
\email{shabbar.raza@fuuast.edu.pk}
\affiliation{Department of Physics, Federal Urdu University of Arts, Science and Technology, Karachi 75300, Pakistan}


\begin{abstract}
\color{black}
With the grand desert hypothesis, we investigate the Supersymmetric (SUSY) Grand Unified Theories (GUTs) with the minimal anomaly mediated supersymmetry breaking (AMSB) as a continuation of our program of studying SUSY GUTs at the future 100 TeV proton-proton Collider such as $\rm FCC_{hh}$ and SppC as well as proton-decay experiment such as Hyper-Kamiokande experiment. We show that for the GUT scale $M_{GUT} \ge 0.84 \times 10^{16}$ GeV, the upper mass bounds on the gluino and first two generation squarks are 8 TeV and 6 TeV, respectively. Therefore, the gluino and first two generation squarks can be probed at the $\rm FCC_{hh}$ and SppC. In addition, we discuss the mass spectra for the colored sparticles, sleptons, pseudo-scalar Higgs and electroweakinos. We also shed light on the possible search strategies of gluino and winos at the future 100 TeV colliders. Moreover, for $\rm M_{\rm GUT} \le  0.84 \times 10^{16}$ GeV, the proton lifetime for dimension-six proton decay via heavy gauge boson exchange is less than or around $5 \times 10^{34}$ years, and thus we can probe it at the Hyper-Kamiokande experiment. To be concrete, we present a couple of benchmark points representing the above two mentioned GUT conditions on $\rm M_{\rm GUT}$.

\color{black}
\end{abstract}
\maketitle

\section{Introduction} 

Supersymmetry (SUSY) solves the gauge hierarchy problem, and provides a very promising extension of the Standard Model (SM). One of the hallmarks of SUSY is providing the gauge coupling unification of the electromagnetic, weak and the strong interactions at some high energy scale~\cite{gaugeunification}.  Under suitable symmetry such as $R$-parity, the lighest supersymmetric particle (LSP) acts a viable dark matter (DM) candidates in the Suppersymmetric Standard Models (SSMs)~\cite{Jungman:1995df}. Moreover, the unification of the gauge couplings strongly suggest the Grand Unified Theories (GUTs)~\cite{Georgi:1974sy,Pati:1974yy,Fritzsch:1974nn,Georgi:1974my}. 

The discovery of SUSY has been one of the most important targets of the Large Hadron Collider (LHC). Although the LHC has not observed any superpartnes in Run-1, Run-2 and Run-3, but this situation does not reduce the motivation for SUSY hunting in general and SUSY GUTs in particular. The accumulation of data from the LHC runs has put stringent bounds on masses of the SUSY particles. For example,  the lower mass bounds on gluino, 
the first two generation of squarks, stop, and sbottom are  2.3 TeV, 1.9 TeV, 1.25 TeV, and 1.5 TeV, 
respectively~\cite{ATLAS-SUSY-Search, Aad:2020sgw, Aad:2019pfy, CMS-SUSY-Search-I, CMS-SUSY-Search-II}. 
This situation suggests that even if we discover SUSY, we may have to confront the EW fine-tuning problem (EWFT). But one can find some promising solutions such as ~\cite{Dimopoulos:1995mi, Cohen:1996vb, Kitano:2006gv, LeCompte:2011cn, Fan:2011yu, Kribs:2012gx, 
Baer:2012mv, Ding:2015epa,Batell:2015fma,Leggett:2014hha,Du:2015una,Li:2015dil,Ellis:1986yg,Barbieri:1987fn}.  

One of the important features of such natural SSMs is that they generically predict relatively light superpartners spectrum, for
instance, bino, Higgsino, stop, gluino, and sleptons, etc. There are also proposals to explore the new physics beyond the SM, at the future proton-proton (pp) colliders, for instance, 
the ${\rm FCC}_{\rm hh}$~\cite{Benedikt:2018csr}  and SppC~\cite{CEPC-SPPCStudyGroup:2015csa}. The simplest question can be asked, if it is possible to probe SUSY at ${\rm FCC}_{\rm hh}$ and SppC. The answer is no since the sparticles can in principle be very heavy. Interestingly, one can further ask if it is possible to probe SUSY GUTs at the $\rm FCC_{hh}$ and SppC despite facing the EWFT problem. The answer is surprisingly yes if we also consider the future proton decay experiment such as Hyper-Kamiokande experiment. The SUSY searches at the 100~TeV pp colliders have been reported previously~\cite{Benedikt:2018csr,CEPC-SPPCStudyGroup:2015csa, Cohen:2013xda, Arkani-Hamed:2015vfh,
Fan:2017rse, Golling:2016gvc}. For the integrated luminosity 30~${\rm ab}^{-1}$,
Wino via Bino decay, gluino ${\tilde g}$ via heavy flavor decay, gluino via light flavor decay,
first-two generation squarks ${\tilde q}$, and stop can be discovered for their masses
up to about 6.5~TeV, 11~TeV, 17~TeV, 14~TeV, and 11~TeV, respectively.
Moreover, if the gluino and first-two generation squark masses are similar,
they can be probed up to 20 TeV.

It is interesting to note that the correlations between the low energy SUSY spectra
and the GUT scale have been studied via the one-loop renormalization group
equations before, and it was found that the bound from dimension-six proton decay
already excludes the gluinos and Winos heavier than about 120 TeV and 40 TeV, respectively,
if their mass ratio $M_3/M_2$ is about 3~\cite{Pokorski:2017ueo}.
Also, we have proposed the concrete scientific goal for the ${\rm FCC}_{\rm hh}$, SppC, and Hyper-Kamiokande experiments to
  probe the supersymmetric GUTs.
 In a previous study~\cite{Ahmed:2020qdt}, with the grand desert hypothesis, it is shown that 
the supersymmetric GUTs can be probed at the  ${\rm FCC}_{\rm hh}$, SppC, and Hyper-Kamiokande experiment. For the GUTs with the GUT scale $M_{GUT} \le 1.0\times 10^{16}$~GeV, the dimension-six proton decay via heavy gauge boson exchange can be probed
at the Hyper-Kamiokande experiment. Also, for the GUTs with 
$M_{GUT} \ge 1.0\times 10^{16}$~GeV, we showed that the GUTs with anomaly and gauge mediated supersymmetry breakings are well within the reaches of the future 100 TeV pp colliders such as the ${\rm FCC}_{\rm hh}$ and SppC, and the supersymmetric GUTs with gravity mediated supersymmetry breaking 
can be probed at the future 160 TeV pp collider. Moreover, we studied the gravity mediation~\cite{Ahmed:2022fjw} and gauge mediation~\cite{Ahmed:2026sax} in details.

In this article, in continuation of this program, with the grand desert hypothesis, we study the minimal 
anomaly mediated supersymmetry breaking (AMSB) realization with the GUT scale conditions $\rm {M_{GUT}\ge 0.84 \times 10^{16}}$ GeV and $\rm {M_{GUT}\le \times 10^{16}}$ GeV chosen so that, within a common reference normalization for dimension-six gauge-boson-mediated proton decay, it corresponds to the benchmark partial lifetimes  $5\times10^{34}$ yr. We show that the viable parameter space with $\rm {M_{GUT}\ge 0.84 \times 10^{16}}$ GeV can be probed at the future 100 TeV proton-proton colliders (definitely depending upon other technical details), and the viable parameter space with $\rm {M_{GUT}\le 0.84 \times 10^{16}}$ GeV is within the reach of the future Hyper-Kamiokande experiment.

The resulting phenomenology has two experimentally distinct components. First, the colored sector contains gluinos and squarks in the multi-TeV regime. Their discovery potential depends on whether the gluino decays dominantly through light-flavor jets, third-generation quarks, or longer cascades through electroweakinos. Second, the anomaly-mediated gaugino relation produces a wino-like lightest neutralino nearly degenerate with the lighter chargino. The small charged-neutral mass splitting suppresses visible decay energy and motivates disappearing-track and soft-pion searches rather than conventional hard-lepton analyses. We therefore treat the gluino and wino signals as correlated parts of one test of the model, not as independent simplified-model targets.

This paper is organized as follows. Section~II summarizes the anomaly-mediated soft terms and the compressed wino sector. Section~III describes the scan, low-energy constraints, and the proton-decay calibration of the nested color samples. Section~IV presents the figure-by-figure spectrum analysis. Section~V develops the gluino and wino search program at hadron and lepton colliders, and Sec.~VI summarizes the conclusions and the theoretical qualifications of the interpretation.

\section{Probing the Supersymmetric Grand Unified Theories with Anomaly Mediation at the Future Proton-Proton Colliders }\label{sec:scan}

Anomaly mediation arises when the visible sector is sequestered from the hidden supersymmetry-breaking sector, so that the dominant visible-sector soft terms are induced by the super-Weyl anomaly\cite{Randall:1998uk,Giudice:1998xp,Luty:2001zv}.  The resulting soft terms are ultraviolet insensitive: they depend on the low-energy beta functions and anomalous dimensions rather than on arbitrary high-scale flavor structures.  This property is one of the main theoretical virtues of AMSB because it naturally suppresses dangerous flavor-changing effects.

The gaugino masses are given by
\begin{equation}
M_a = \frac{\beta_{g_a}}{g_a}\,\mgrav,
\label{eq:amsb_gaugino_general}
\end{equation}
where \(a=1,2,3\) labels the gauge groups \(U(1)_Y\), \(SU(2)_L\), and \(SU(3)_C\), and \(\beta_{g_a}\) is the beta function of the corresponding gauge coupling.  At one-loop order,
\begin{equation}
M_a = \frac{b_a g_a^2}{16\pi^2}\,\mgrav .
\label{eq:amsb_gaugino}
\end{equation}
This equation immediately explains why a gravitino mass of order \(100\)--\(1000\tev\) can generate gaugino masses at the TeV scale: the visible soft masses are loop-suppressed relative to \(\mgrav\).  It also fixes the relative gaugino hierarchy.  Since the \(SU(2)_L\) coefficient yields the smallest absolute weak-scale gaugino mass in the usual AMSB spectrum, the lightest neutralino and lighter chargino are typically wino-like.

The trilinear parameters are similarly determined by the Yukawa beta functions,
\begin{equation}
A_y = -\frac{\beta_y}{y}\,\mgrav,
\label{eq:amsb_A}
\end{equation}
while the scalar masses follow from the anomalous dimensions,
\begin{equation}
m_i^2 = -\frac{1}{4}\left(
\frac{\partial \gamma_i}{\partial g_a}\beta_{g_a}
+
\frac{\partial \gamma_i}{\partial y}\beta_y
\right)\mgrav^2 .
\label{eq:amsb_scalar}
\end{equation}
Here \(\gamma_i\) is the anomalous dimension of the chiral superfield \(i\).  Equations~\eqref{eq:amsb_gaugino_general}--\eqref{eq:amsb_scalar} are the reason AMSB is predictive: the soft spectrum is tied to the renormalization-group structure of the MSSM.

\subsection{Tachyonic sleptons and phenomenological AMSB}

Pure AMSB predicts negative mass squared values for sleptons in the minimal MSSM.  This tachyonic slepton problem is a well-known obstruction to the most minimal version of the theory\cite{Pomarol:1999ie,Gherghetta:1999sw}.  A simple phenomenological solution is to add a universal positive scalar mass contribution,
\begin{equation}
m_i^2 \rightarrow m_i^2 + \mzero^2 .
\label{eq:m0_shift}
\end{equation}
This gives the phenomenological AMSB or minimal AMSB parameter set
\begin{equation}
\mzero, \qquad \mgrav, \qquad \tanb, \qquad {\rm sign}(\mu).
\label{eq:amsb_inputs}
\end{equation}
The gravitino mass \(\mgrav\) controls the anomaly-mediated gaugino and trilinear mass scale, while \(\mzero\) controls the scalar-mass shift needed to make the sleptons physical.  The parameter \(\tanb\) mainly affects the third-generation scalar and Higgs sectors through bottom and tau Yukawa interactions.

A robust phenomenological consequence of AMSB is the compressed wino-like electroweakino sector,
\begin{equation}
\mchar \simeq \mchi .
\label{eq:wino_degeneracy}
\end{equation}
At tree level the charged and neutral winos are nearly degenerate, and the physical splitting is generated by electroweak radiative corrections.  This can make the chargino long-lived on detector scales, leading to disappearing-track signatures.  Thus AMSB has a distinctive collider fingerprint: conventional jets plus missing-energy searches probe the colored sector, while disappearing-track and compressed-spectrum analyses probe the electroweakino sector.


\section{Scanning Procedure, Constraints, and Proton-Decay Calibration}
\label{sec:numerical}
\color{black}
We use \textsc{ISAJET}~7.85~\cite{ISAJET} to explore the phenomenological minimal-AMSB parameter space. The weak-scale gauge and third-generation Yukawa couplings are evolved to the unification scale with the MSSM renormalization-group equations in the $\overline{\mathrm{DR}}$ scheme. The scale $M_{\rm GUT}$ is defined by $g_1(M_{\rm GUT})=g_2(M_{\rm GUT})\equiv g_G$. To parameterize unknown GUT-scale threshold effects, we allow the strong coupling mismatch
to remain within $3\%$ in magnitude~\cite{Hisano:1992jj}. The soft terms and dimensionless couplings are iterated between $M_Z$ and $M_{\rm GUT}$ using the full two-loop RGEs until numerical convergence is reached. Supersymmetric threshold effects are included at $M_{\rm SUSY}$
and the RGE-improved one-loop Higgs potential is minimized at the optimized supersymmetry scale~\cite{Pierce:1996zz}.

The Metropolis--Hastings scan~\cite{Belanger:2009ti,Baer:2008jn} covers
\begin{align}
1\tev &\leq m_0\leq10\tev,\nonumber\\
100\tev &\leq m_{3/2}\leq1200\tev,\nonumber\\
2&\leq\tan\beta\leq60,
\label{eq:amsb_scan_ranges}
\end{align}
with $\mu>0$, $m_t=173.3\,\mathrm{GeV}$, and the \textsc{ISAJET} input $m_b^{\overline{\mathrm{DR}}}(M_Z)=2.83\,\mathrm{GeV}$. Every retained point satisfies radiative electroweak symmetry breaking. The yellow sample additionally satisfies
\begin{align}
122\gev &\leq m_h\leq128\gev,\nonumber\\
m_{\widetilde g}&\geq2.2\tev,\nonumber\\
0.8\times10^{-9}&\leq {\rm BR}(B_s\to\mu^+\mu^-)
\leq6.2\times10^{-9},\nonumber\\
2.99\times10^{-4}&\leq {\rm BR}(b\to s\gamma)
\leq3.87\times10^{-4},\nonumber\\
0.15&\leq
\frac{{\rm BR}(B_u\to\tau\nu)_{\rm AMSB}}
{{\rm BR}(B_u\to\tau\nu)_{\rm SM}}
\leq2.41,
\label{eq:basic_constraints}
\end{align}
together with the remaining sparticle limits implemented in the scan~\cite{ATLAS:2012yve,CMS:2012qbp,Agashe:2014kda,Aaij:2012nna,Amhis:2012bh,Asner:2010qj}.

\subsection{Dimension-six proton decay and the nested GUT selections}

Heavy $X$ and $Y$ gauge-boson exchange induces the dimension-six mode $p\to e^+\pi^0$. For the purpose of organizing the scan, we use the reference scaling
\begin{equation}
{\tau(p\to e^+\pi^0)}
\simeq10^{35}\ {\rm yr}
\left(\frac{M_X}{10^{16}\gev}\right)^4,
\label{eq:proton_reference}
\end{equation}
where the normalization represents fixed reference values of the unified coupling, short- and long-distance renormalization factors, flavor coefficients, and hadronic matrix elements. Identifying $M_X\simeq M_{\rm GUT}$ then gives
\begin{align}
M_{\rm GUT}=0.84\times10^{16}\gev
&\Longleftrightarrow \tau\simeq5\times10^{34}\ {\rm yr}.
\label{eq:gut_lifetime_map}
\end{align}
These are benchmark translations, not universal model-independent mass limits. GUT thresholds, the ratio $M_X/M_{\rm GUT}$, $\alpha_G$, flavor rotations, and the lattice matrix element can shift the inferred lifetime. The current Super-Kamiokande limit and the projected Hyper-Kamiokande sensitivity nevertheless motivate presenting the results directly in lifetime-ordered subsets~\cite{SuperK:2020proton,HyperK:2018design}. We have employed the $M_{\rm GUT}=0.84\times10^{16}\gev
\Longleftrightarrow \tau\simeq5\times10^{34}\ {\rm yr}$ as a constraint at $\rm M_{\rm GUT}$.



\section{Numerical Results and Spectrum Analysis}
\label{sec:numerical_results}
\color{black}
The figures display the hierarchy generated by progressively stronger constraints. Gray points satisfy the conditions of successful EWSB and neutralino as the LSP. Yellow solutions form a subset of gray points satisfy the basic constraints such as  Higgs-mass bound, sparticle mass bounds and B-physics bounds as given in eq.~\ref{eq:basic_constraints}. In addition to the basic constraints, yellow solutions also satisfy $\rm{M_{\rm GUT}}\le 0.84 \times 10^{16}$ GeV. Red points form as subset of yellow solutions and consistent with the $\rm GUT$ condition $\rm{M_{\rm GUT}}\ge 0.84 \times 10^{16}$ GeV.

Moreover, in all the plots with the gluino mass on the vertical axis, the horizontal lines at
\begin{equation}
m_{\widetilde g}=\ 11,\ 17\tev
\label{eq:gluino_reach_lines}
\end{equation}
are representative reach markers rather than model-independent exclusions.  The $11$ and $17\,\mathrm{TeV}$ lines illustrate the spread of projected $100\,\mathrm{TeV}$ sensitivities for heavy-flavor and light-flavor gluino topologies, respectively~\cite{Cohen:2013xda,Arkani-Hamed:2015vfh,Fan:2017rse,Golling:2016gvc}. Actual sensitivity depends on squark masses, branching fractions, the neutralino mass, compression, luminosity, detector performance, and systematic uncertainties.

\subsection{Fundamental AMSB parameter space}

\begin{figure*}[t]
\centering
\includegraphics[width=0.48\textwidth]{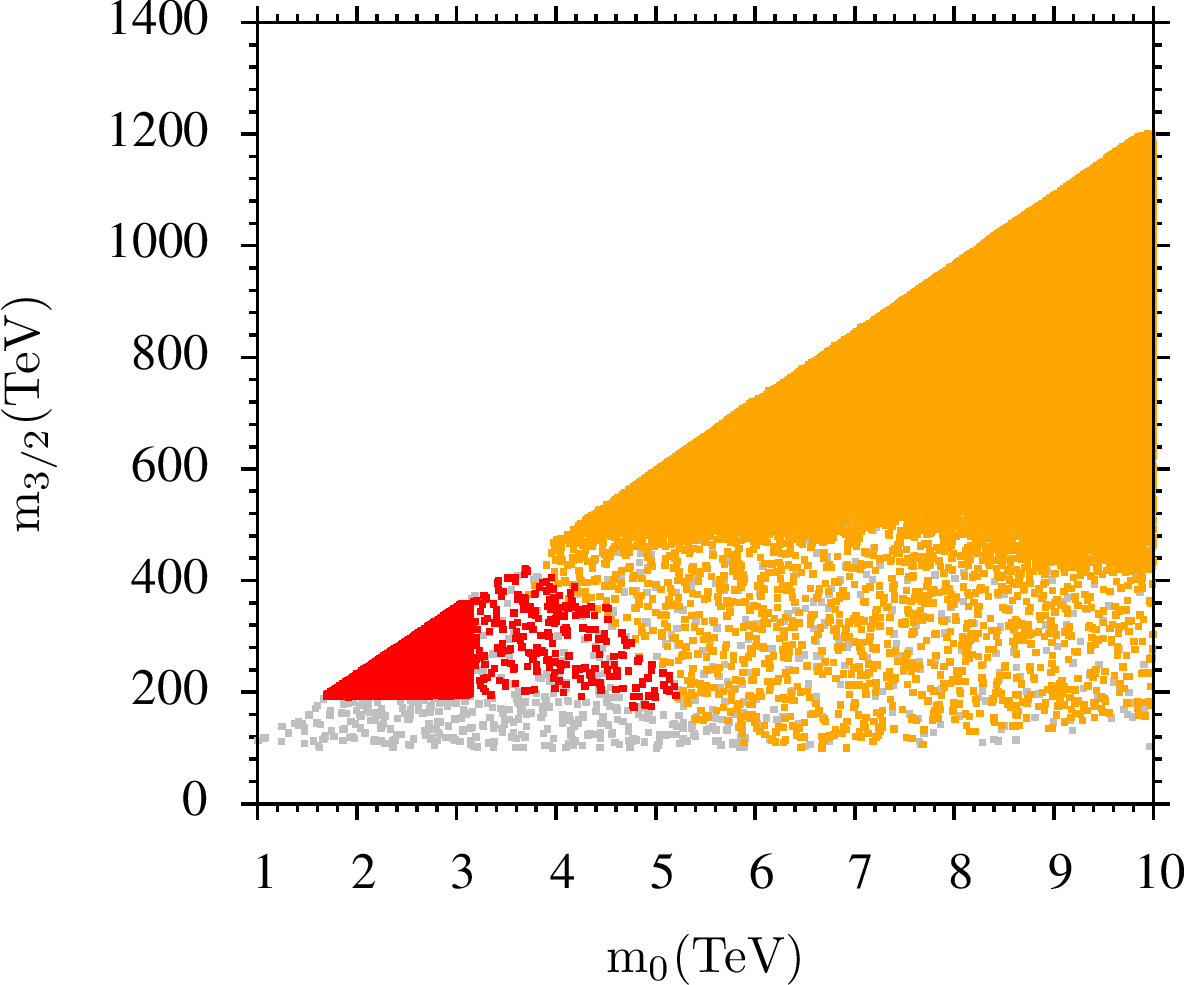}\hfill
\includegraphics[width=0.48\textwidth]{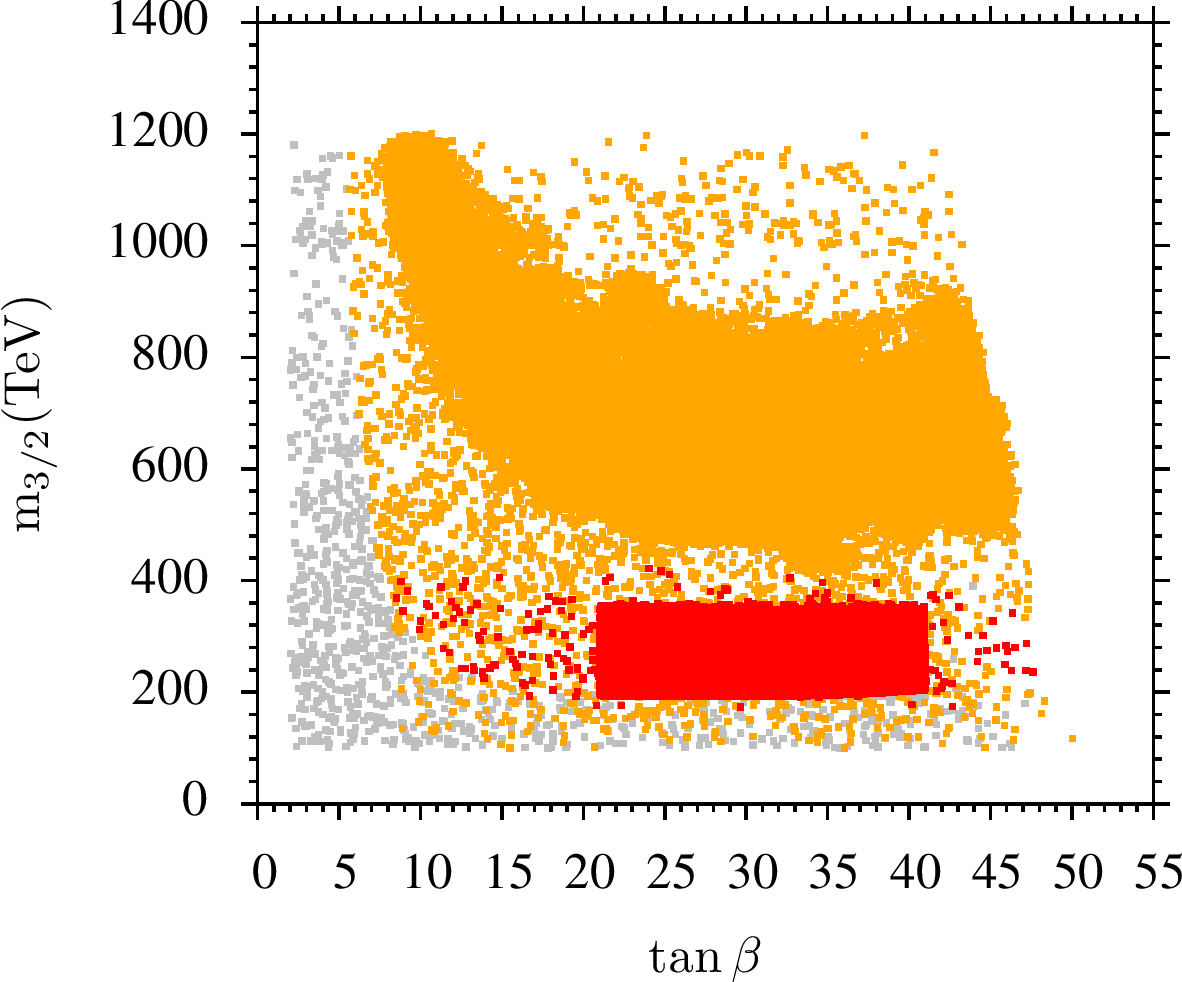}
\caption{Fundamental AMSB parameter space. Left: $m_0$ versus $m_{3/2}$. Right: $\tan\beta$ versus $m_{3/2}$. Gray points denote the raw scan, yellow points satisfy the basic constraints, and red subsets with $M_{\rm GUT} > 0.84 \times 10^{16}$ GeV corresponding to proton-lifetime benchmarks above  $5\times10^{34}$ yr, respectively. The stricter selection progressively localize $m_0$, $m_{3/2}$, and $\tan\beta$.}
\label{fig:fundamental}
\end{figure*}

Figure~\ref{fig:fundamental} shows the $(m_0,m_{3/2})$ and $(\tan\beta,m_{3/2})$ planes. The gray and yellow samples occupy broad regions because $m_0$ and $m_{3/2}$ can compensate each other in the scalar spectrum: a larger positive $m_0^2$ contribution cures the tachyonic-slepton problem over an increasingly wide range of anomaly-mediated masses. The points satisfy $M_{\rm GUT} > 0.84 \times 10^{16}$ GeV condition are shown in the red color. This is a stringent constraint which restrict the red solutions with $m_{0}\sim$ 1.6-5.2 TeV and $m_{3/2} \sim$ 200-400 TeV. 

The $\tan\beta$ panel gives complementary information. The red sample favors moderate-to-large $\tan\beta$, approximately $10\lesssim\tan\beta\lesssim 47$. This correlation arises because $\tan\beta$ modifies the bottom and tau Yukawa contributions, the stau spectrum, and the Higgs potential. Thus the proton-lifetime selection is not merely a cut on a high scale: through RGE evolution it singles out a correlated low-energy island.

\subsection{Gluino mass versus the AMSB inputs}

\begin{figure*}[t]
\centering
\includegraphics[width=0.48\textwidth]{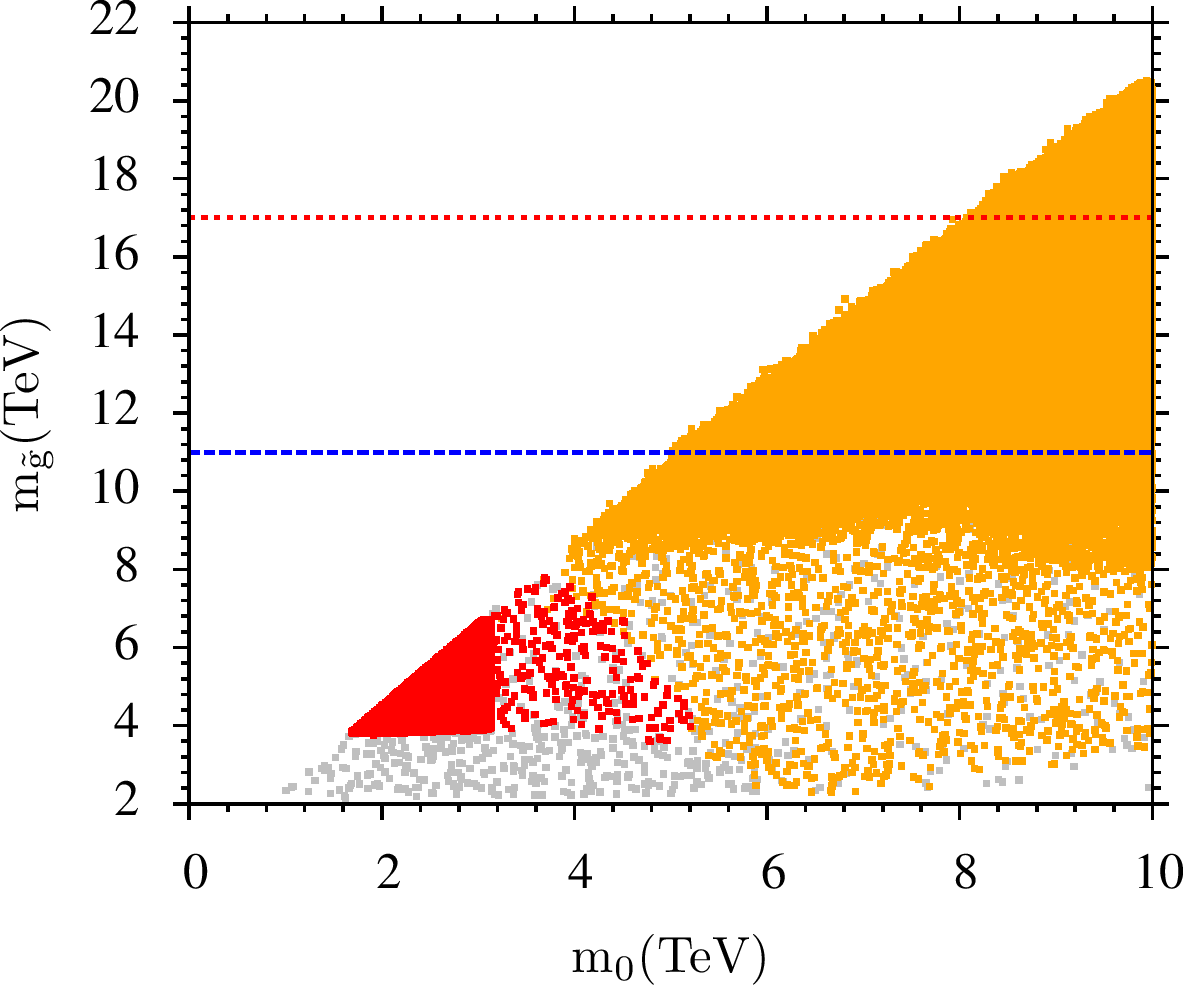}\hfill
\includegraphics[width=0.48\textwidth]{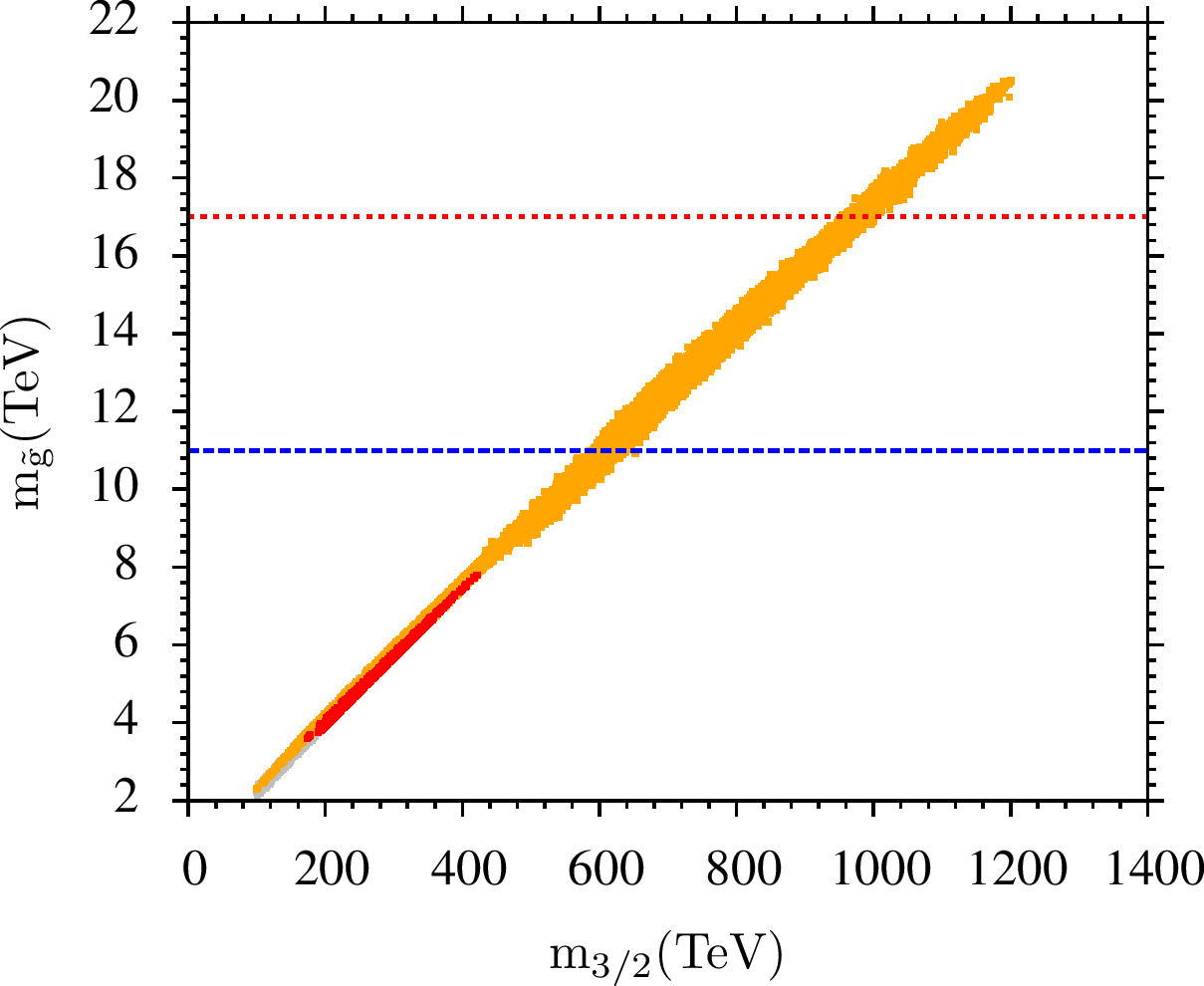} \hfill
\includegraphics[width=0.48\textwidth]{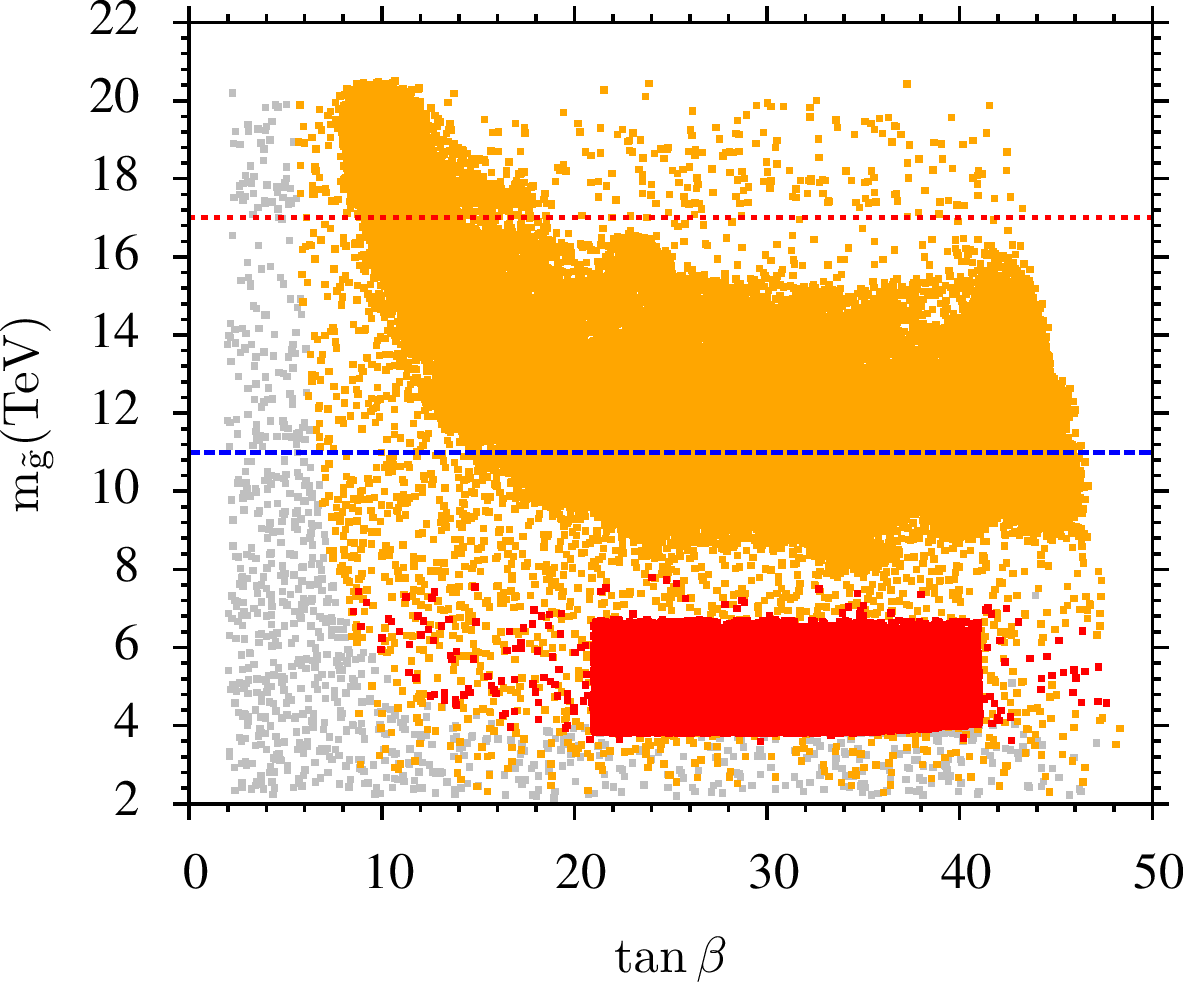} \hfill
\caption{Gluino correlations with the AMSB inputs. Left: $m_{\widetilde g}$ versus $m_0$. Right: $m_{\widetilde g}$ versus $m_{3/2}$. Bottom: $m_{\widetilde g}$ versus $\tan\beta$. The horizontal lines at $11$, and $17\,\mathrm{TeV}$ are topology-dependent future-collider reference sensitivities. The strict red proton-lifetime subset lies predominantly in the several-TeV gluino region.}
\label{fig:gluino_inputs}
\end{figure*}

Figure~\ref{fig:gluino_inputs} is the most direct visualization of the collider implication. In the top panel the basic-constraint sample extends beyond two guide lines, so minimal AMSB without the GUT information can contain a colored sector inaccessible even to a $100\,\mathrm{TeV}$ machine. The GUT selections substantially change this conclusion. The strict red sample is localized at roughly $m_{\widetilde g}\sim4.0$--$8\,\mathrm{TeV}$ corresponding to $m_{0}\sim 1.6-5.2$ TeV and $m_{3/2}\sim 240-420$ TeV. Yellow points extend into and above the marker regions $17\,\mathrm{TeV}$ and $11\,\mathrm{TeV}$.
A $100\,\mathrm{TeV}$ collider should provide a much more robust test.

In the bottom panel, the $m_{\widetilde g}$--$\tan\beta$ plot shows that the same red solutions occupy a restricted horizontal band and a moderate-to-large $\tan\beta$ interval. This narrow target is valuable for experimental optimization: it motivates dedicated analyses near the several-TeV gluino scale rather than a search designed only for an asymptotically heavy spectrum.

\subsection{Colored scalar spectrum}

\begin{figure*}[t]
\centering
\includegraphics[width=0.48\textwidth]{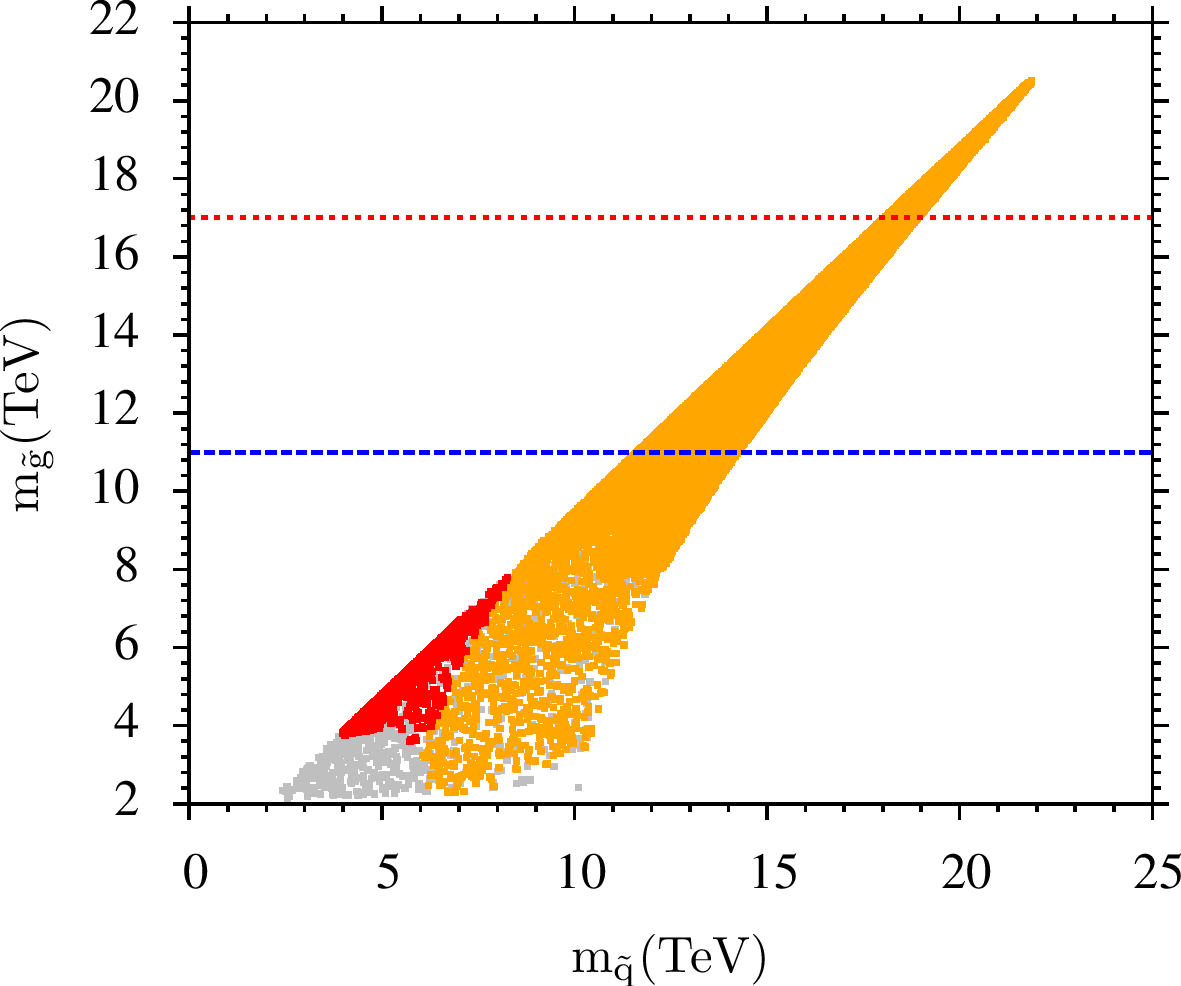}\hfill
\includegraphics[width=0.48\textwidth]{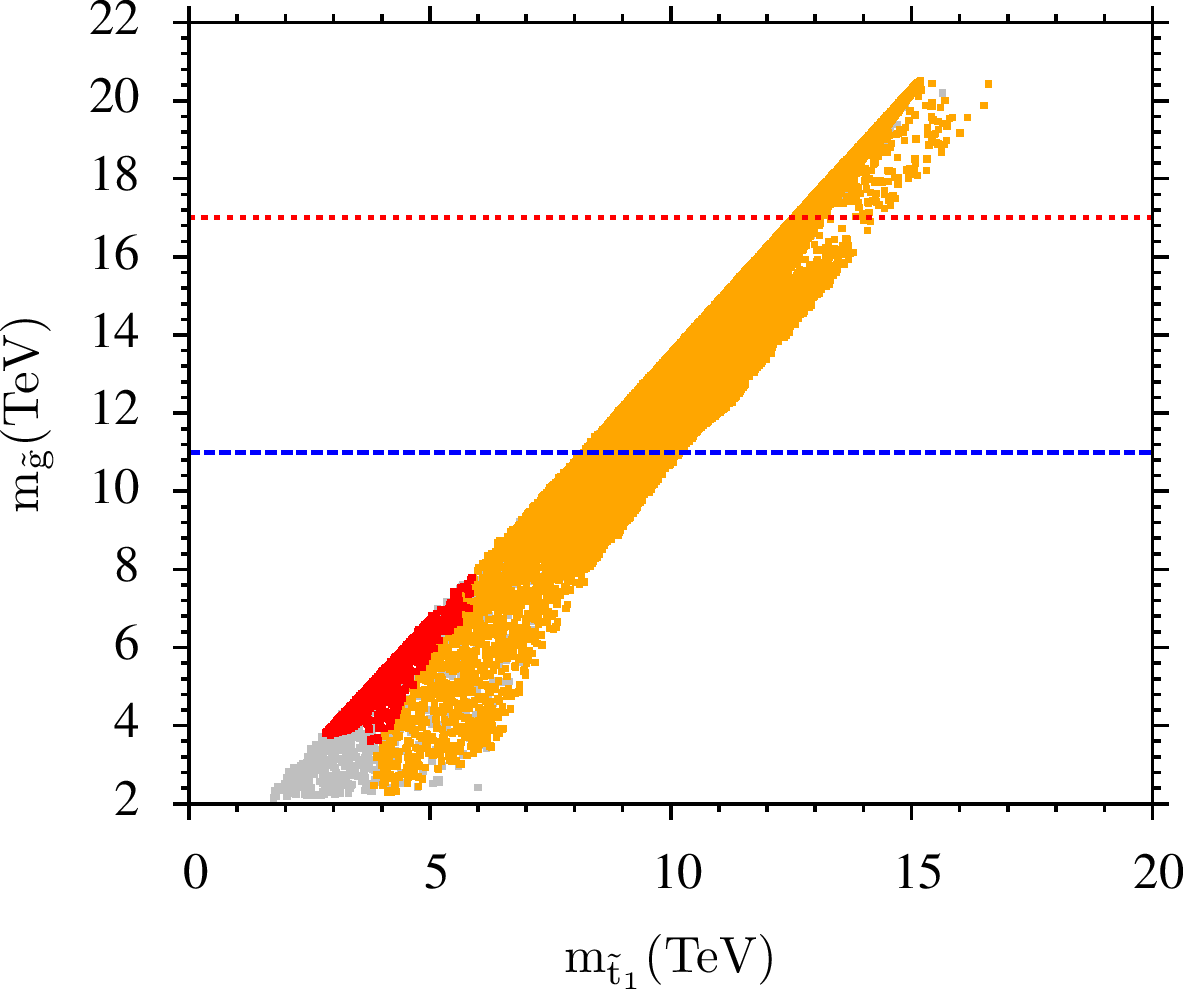}
\caption{Colored scalar spectrum. Left: $m_{\widetilde g}$ versus the first- and second-generation squark mass. Right: $m_{\widetilde g}$ versus $m_{\widetilde t_1}$. The QCD-driven squark--gluino correlation is tighter than the stop--gluino relation because the latter receives top-Yukawa and mixing effects. The strict red subset predicts a correlated several-TeV colored spectrum suitable for combined gluino, squark, and third-generation searches.}
\label{fig:colored}
\end{figure*}

Figure~\ref{fig:colored} shows the first-two-generation squark and lighter-stop correlations with gluino. The near-linear squark--gluino relation follows from the common QCD origin of their running. In AMSB the gluino mass is proportional to the $SU(3)_C$ beta function, while the squark soft masses receive large gluino-driven RGE contributions. The red sample selects squarks of approximately $5$-$8\,\mathrm{TeV}$ together with gluinos near $5$-$7\,\mathrm{TeV}$. The red solutions can be probed at the future 100 TeV colliders such as $\rm{LHC_{\rm hh}}$ and SppC through gluino decays via heavy flavors. When squarks and gluinos are comparable in mass, associated production and the simultaneous contribution of squark and gluino pair production can strengthen a future hadron-collider analysis relative to a decoupled-squark simplified model. The yellow points which satisfy $M_{\rm GUT} < 0.84 \times 10^{16}$ GeV, can be explored at Hyper-Kamiokande experiment. In addition to it, dotted blue and red lines correspond to gluino masses of 11 TeV and 17 TeV which represent gluino mass bound respectively coming from gluino decays through heavy and light flavors, also put upper bounds on the first-two generation squark masses. We can see that the upper bounds due to dotted blue and red lines are 14 TeV and 19 TeV respectively.

The stop correlation is broader because $y_t$, left-right mixing, and electroweak symmetry breaking modify the two stop eigenvalues. The red solutions contain multi-TeV stops, approximately in the $3$-$6\,\mathrm{TeV}$ range in the figure, which help generate the observed Higgs mass. Their proximity to the gluino scale also implies that gluino decays involving virtual or on-shell third-generation squarks can be relevant. This motivates top- and bottom-enriched searches in addition to inclusive light-jet plus missing-momentum channels. Here we make a comment that we noticed that in our scans, for all the red solutions, but for some of the yellow solutions, gluino mass is heavier than the squark masses, otherwise squarks are heavier than gluinos. It is clear from the figure that all the red points are within the reach of future 100 TeV proton-proton colliders such as $\rm {LHC_{hh}}$ and SppC and yellow points which satisfy $\rm M_{\rm GUT} < \times 10^{16}$ GeV can be probed at Hyper-Kamiokande experiments. The bounds from gluino decays through heavy flavor (blue-dotted line) and the light flavors (red-dotted line) restrict the lighter stop to 10 TeV and 14 TeV respectively.  

\subsection{Stau and heavy-Higgs sectors}

\begin{figure*}[t]
\centering
\includegraphics[width=0.48\textwidth]{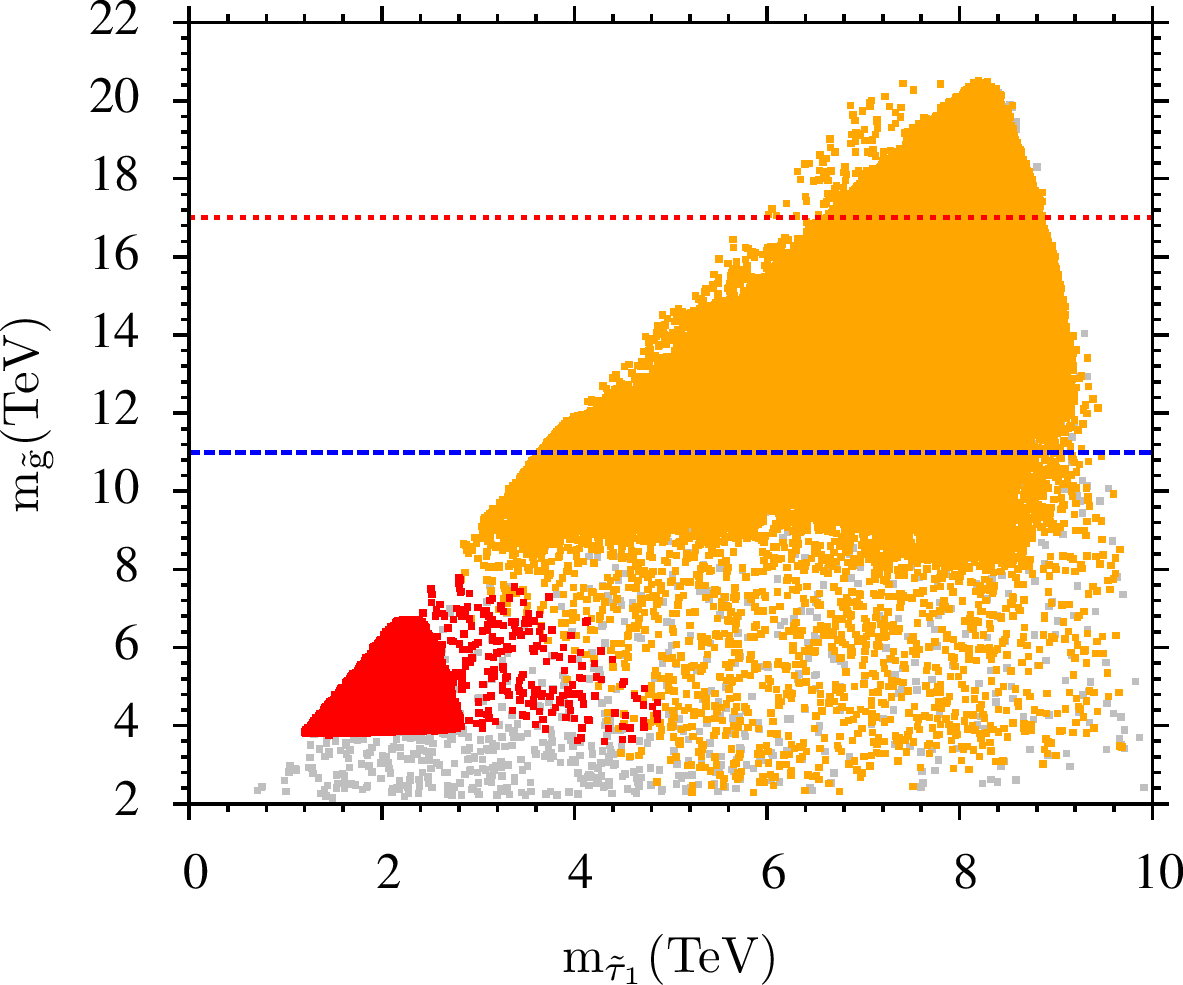}\hfill
\includegraphics[width=0.48\textwidth]{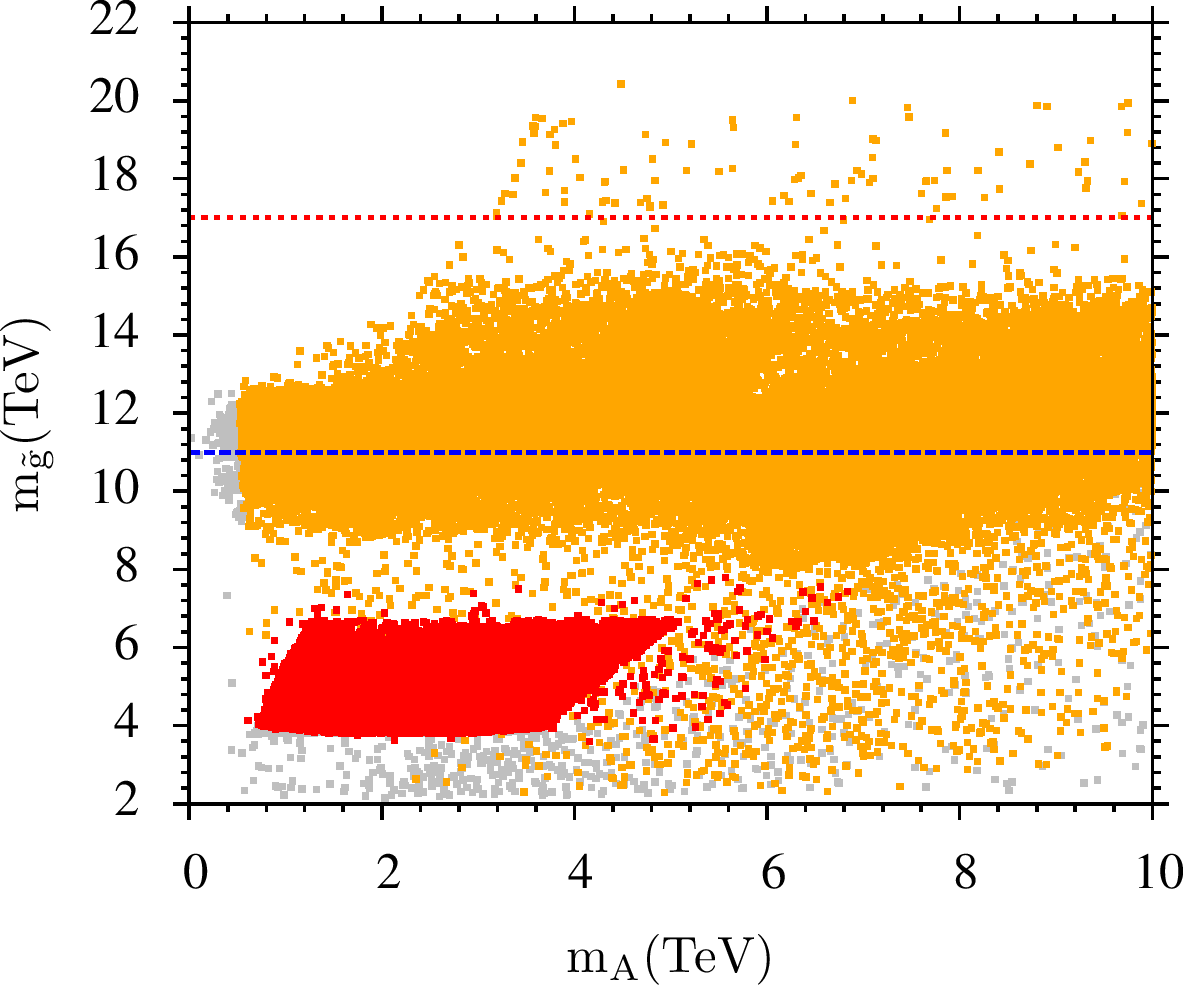}
\caption{Stau and heavy-Higgs sectors. Left: $m_{\widetilde g}$ versus $m_{\widetilde\tau_1}$. Right: $m_{\widetilde g}$ versus $m_A$. The stau mass probes the phenomenological $m_0^2$ shift, while $m_A$ probes the Higgs soft terms and electroweak symmetry breaking.}
\label{fig:stau_higgs}
\end{figure*}

The left panel of Fig.~\ref{fig:stau_higgs} directly tests the positive scalar-mass shift introduced to cure tachyonic sleptons. The red points select a lighter stau at the few-TeV scale while maintaining a gluino near the several-TeV region. This is a nontrivial consistency relation: the value of $m_0$ must be large enough to render the sleptons physical but not so large that the entire scalar sector decouples. We see that for the red points, lighter stau mass is in the range 1.2 to 4.8 TeV. These solutions are expected to be probed at the future 100 TeV colliders. The yellow solutions can be explored at the Hyper-Kamiokande experiment. The blue and the red dotted lines put the upper bounds on the lighter stau mass of 9.6 TeV and 8.8 TeV. 

The pseudoscalar-Higgs panel is broader because $m_A$ depends on both Higgs soft masses, $\mu$, and the electroweak symmetry-breaking conditions. The red solutions shown occupy roughly the $m_A\sim 0.4$-$6.8\,\mathrm{TeV}$ range. Direct heavy-Higgs sensitivity will be strongly $\tan\beta$ dependent, but the $m_A$ correlation remains useful as a consistency test if a colored or wino signal is found. It is evident from the plot that $m_{A}$ is insensitive of gluino mass.

\subsection{Right- and left-handed sleptons}

\begin{figure*}[t]
\centering
\includegraphics[width=0.48\textwidth]{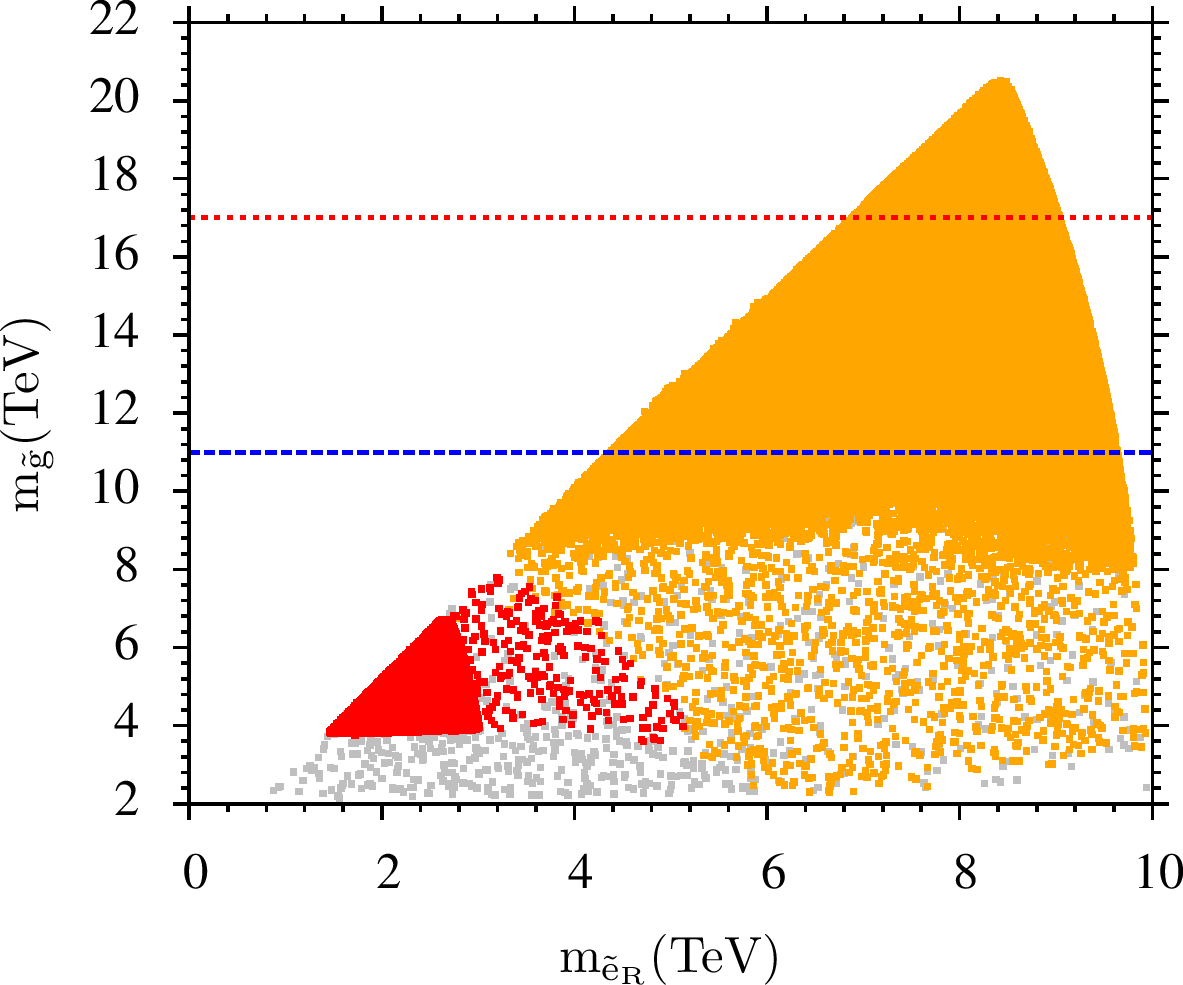}\hfill
\includegraphics[width=0.48\textwidth]{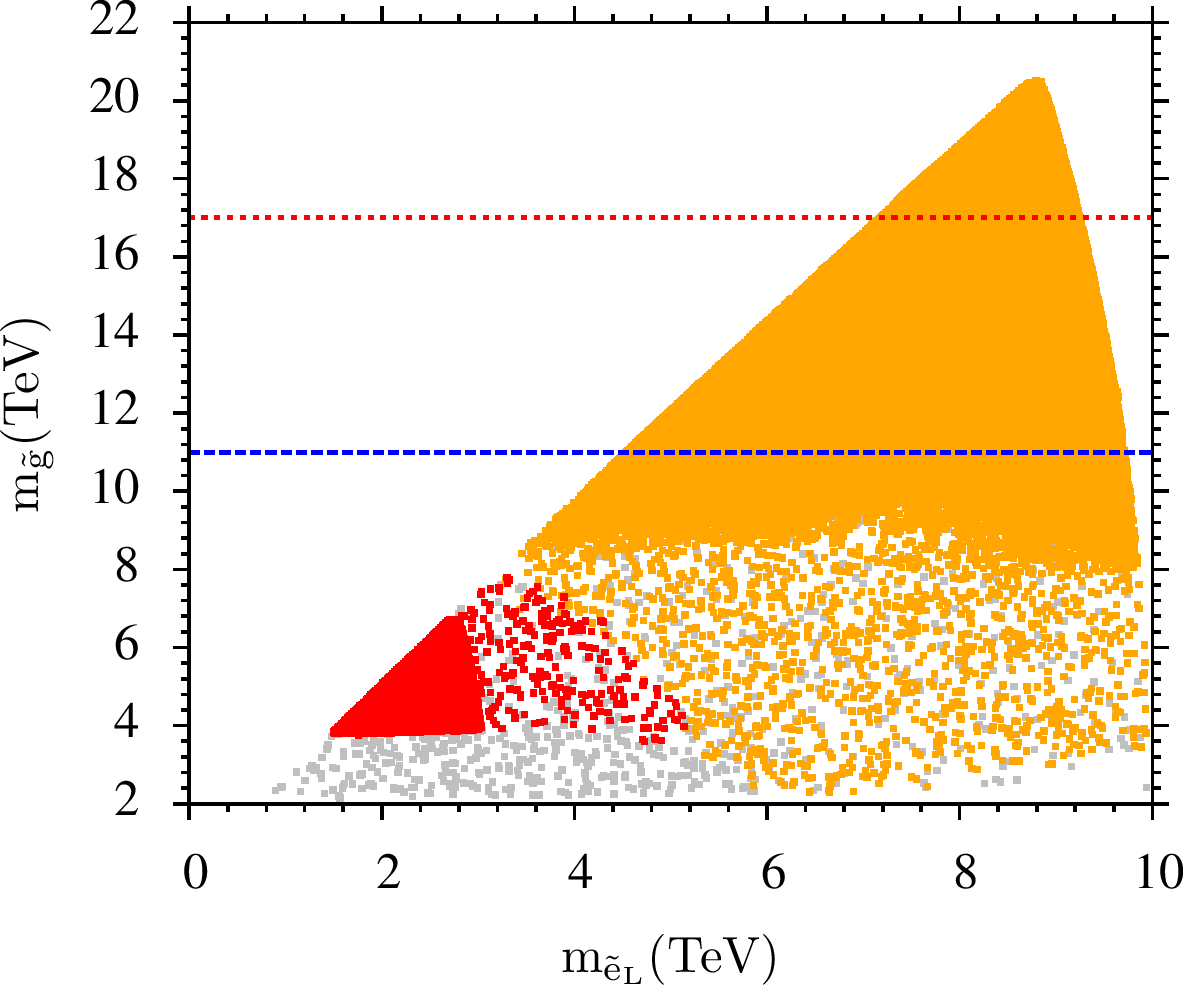}
\caption{Right- and left-handed selectron sectors. Left: $m_{\widetilde g}$ versus $m_{\widetilde e_R}$. Right: $m_{\widetilde g}$ versus $m_{\widetilde e_L}$. The nested GUT selections correlate the slepton mass shift with the gluino and wino scales.}
\label{fig:sleptons}
\end{figure*}

Figure~\ref{fig:sleptons} shows that the red sample selects right- and left-handed sleptons in the few-TeV region. The right-handed states are controlled mainly by hypercharge and the universal shift, whereas left-handed sleptons also receive the $SU(2)_L$ contribution. Their similar but nonidentical bands provide a direct test of the scalar deformation used to solve the pure-AMSB tachyonic-slepton problem. Although direct production may be difficult, the slepton spectrum can affect electroweakino decay amplitudes and is an important consistency condition for reconstructing the mediation mechanism. In both the plots red points are mass range of 1.6 to 5.2 TeV while the blue-dotted and red-dotted lines put the upper mass bound on the slepton masses of 9.6 and 8.2 TeV respectively.

\subsection{Compressed wino-like electroweakinos}

\begin{figure*}[t]
\centering
\includegraphics[width=0.48\textwidth]{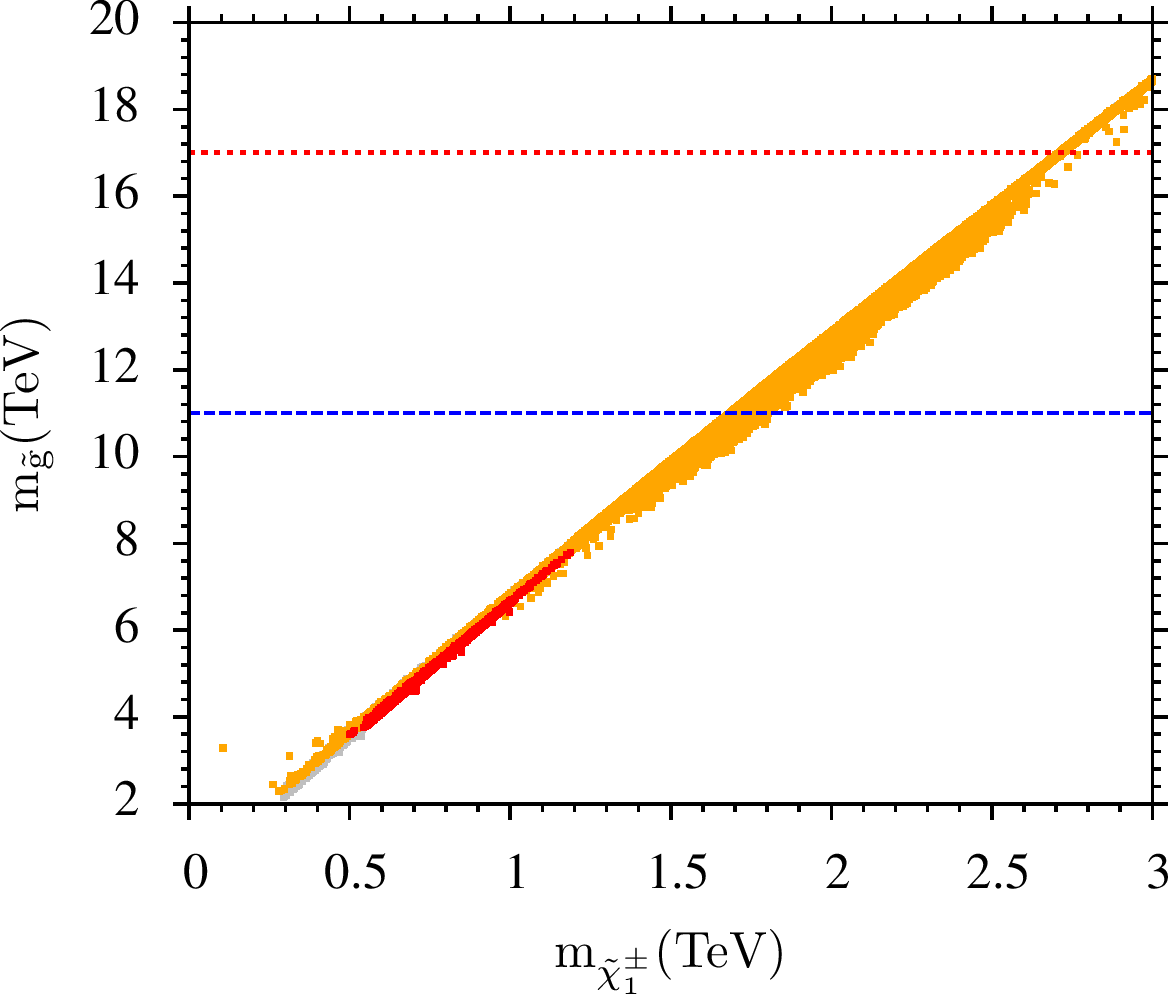}\hfill
\includegraphics[width=0.48\textwidth]{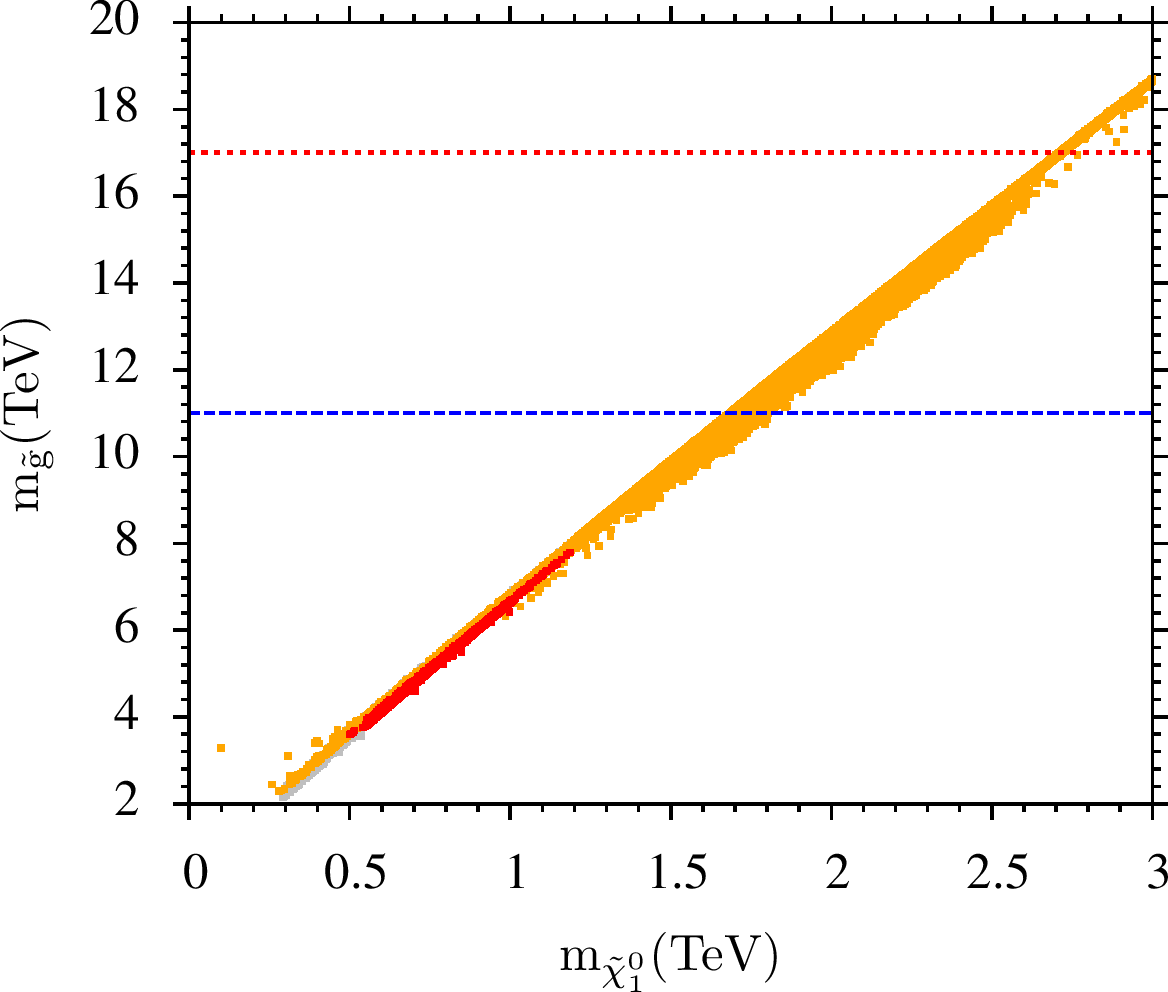}
\caption{Electroweakino correlations. Left: $m_{\widetilde g}$ versus $m_{\widetilde\chi_1^\pm}$. Right: $m_{\widetilde g}$ versus $m_{\widetilde\chi_1^0}$. The almost coincident chargino and neutralino bands demonstrate the compressed wino-like spectrum. The strict red GUT subset predicts sub-TeV to approximately TeV winos correlated with several-TeV gluinos.}
\label{fig:ewkino}
\end{figure*}

Figure~\ref{fig:ewkino} displays the cleanest AMSB fingerprint. The neutralino and chargino panels are nearly identical because $\widetilde\chi_1^0$ and $\widetilde\chi_1^\pm$ are dominantly the neutral and charged components of the same $SU(2)_L$ wino multiplet. Their masses scale with $m_{3/2}$ through $M_2=(\beta_{g_2}/g_2)m_{3/2}$, while the gluino follows the corresponding $SU(3)_C$ coefficient. The resulting straight bands are therefore a direct graphical test of the anomaly-mediated gaugino-mass relation.

The strict red subset corresponds approximately to $m_{\widetilde\chi_1^\pm},m_{\widetilde\chi_1^0}\sim0.7$--$1.1\,\mathrm{TeV}$. The red and blue dotted lines also show upper mass bounds on the chargino and neutralino as 1.7 TeV 2.7 TeV. These ranges should not be interpreted with ordinary hard-lepton electroweakino limits: the visible pion from $\widetilde\chi_1^\pm\to\widetilde\chi_1^0\pi^\pm$ is soft, and the chargino may leave only a short track segment. The relevant observables are the charged-neutral splitting, proper lifetime, track length, boost, and ISR activity.

\section{Collider Interpretation and Search Strategy}
\label{sec:collider_strategy}
\color{black}
\subsection{Gluino searches at a $100\,\mathrm{TeV}$ collider}

The gluino is a Majorana color octet and therefore has a rapidly increasing pair-production rate with hadron-collider energy. In the AMSB spectra shown here, its decay pattern is controlled by the relative squark masses and by the compressed wino endpoint. Three complementary classes of analyses are required.

First, if first- and second-generation squarks mediate the dominant decay, one expects
\begin{equation}
pp\to\widetilde g\widetilde g,
\qquad
\widetilde g\to q\bar q\,\widetilde\chi_{1,2}^0
\ \text{or}\ q\bar q'\,\widetilde\chi_1^\pm,
\end{equation}
leading to high-multiplicity jets plus missing transverse momentum. This topology motivates the larger, light-flavor-like reach represented by the $17\,\mathrm{TeV}$ line. Second, when stop and sbottom exchange is important, the final state contains $t\bar t$, $b\bar b$, or $tb$ together with winos. Searches using multiple $b$ tags, reconstructed boosted tops, same-sign dileptons, and large effective mass are then central; this more difficult topology motivates the lower $11\,\mathrm{TeV}$ reference. Third, spectra with reduced visible energy require ISR-assisted and shape-based analyses in $H_T$, $M_{\rm eff}$, $E_T^{\rm miss}$, and recursive-jigsaw or related kinematic variables.

\subsection{Wino searches at the LHC and future proton colliders}

Direct wino production proceeds through electroweak Drell--Yan processes,
\begin{equation}
pp\to\widetilde\chi_1^+\widetilde\chi_1^-,\qquad
pp\to\widetilde\chi_1^\pm\widetilde\chi_1^0,
\end{equation}
but the compressed decay releases too little visible energy for standard triggers. A hard ISR jet or photon supplies the trigger and recoil, while the signal is identified by one or more disappearing tracks. Dedicated recovery of the soft pion from the chargino decay can improve sensitivity to shorter tracks. Current ATLAS and CMS searches already use short-track, energy-loss, jet, $b$-jet, lepton, and soft-track information~\cite{ATLAS:2026disappearing,CMS:2024disappearing}. Their limits are strongly lifetime dependent; consequently, the lower part of the red mass interval cannot be declared excluded or allowed without computing the point-by-point wino purity, $\Delta m_{\widetilde W}$, and lifetime.

At higher energies, the disappearing-track channel is substantially more powerful than a pure mono-jet analysis. Studies of a $100\,\mathrm{TeV}$ collider find sensitivity extending into the multi-TeV wino regime, with the precise reach controlled by the innermost tracker radius, track-hit requirements, pileup, and background extrapolation~\cite{Han:2018wimps,Saito:2019rtg,Grilli:2014iwa}. 

A decisive analysis should combine direct and cascade production. Direct Drell--Yan events constrain the electroweak quantum numbers with minimal dependence on the colored spectrum. Gluino and squark cascades can yield more boosted charginos and hard trigger objects, increasing track acceptance. Observation of both channels with a common reconstructed chargino lifetime would strongly support the AMSB mass relation.

\subsection{Benchmark spectra}
For the convenience of the readers, we have added two benchmark points: point 1 belongs to the red solutions and point 2 belongs to the yellow solutions. For point 1 $\rm M_{\rm GUT}=0.964 \times 10^{16}$ GeV and for point 2 $\rm M_{\rm GUT}=0.677 \times 10^{16}$ GeV. Here we see that for both the points the color sparticles are in multiple TeV range. Moreover the electroweakinos have the typical AMSB spectrum reflecting pure winos with almost degenerate neutralinos and charginos. For point 1 the mass splitting $\Delta m_{\widetilde\chi^0_{\pm}-\widetilde\chi^0_{1}}=$ 173.706 MeV and for point 2 the mass splitting $\Delta m_{\widetilde\chi^{\pm}_{1}-\widetilde\chi^0_{1}}=$ 174.560 MeV. With such small mass splitting, the dominant decay channel for charge winos will be $\tilde \chi^{\pm}_{1} \rightarrow \tilde\chi^0_{1} \pi^{\pm} $  with estimated  [rpper decay length ($c\tau \approx$) of 4.322 cm and 4.218 cm respectively, indicating the standard ABSB disappearing track configuration. We also notice that at the 100 TeV proton-proton colliders such $\rm LHC_{\rm hh}$ and SppC, there is a good opportunity to prob such a scenario via gluino-pair production, direct wino production and disappearing tracks embedded in thrid-generation squark cascades. One can potentially explore wino LSP, resolve flavor structure, reconstruct heavier electroweakinos and probe AMSB mass relations.

\begin{table}[t]
\centering
\scalebox{1.0}{
\begin{tabular}{lcc}
\hline\hline
 & Point 1 & Point 2\\
\hline
$m_0$ &  2742.7 & 8021.90 \\
$m_{3/2}$& 246780    & 668280 \\
$\tan\beta$& 21.07 & 42.43 \\
$M_{\rm GUT}$&$0.964\times10^{16}$&$0.677\times10^{16}$\\
\hline
$m_h$&123&125\\
$m_H$& 3932& 2449 \\
$m_A$& 3906 & 2433\\
$m_{H^\pm}$&3932 & 2451\\
$m_{\widetilde\chi^0_{1,2}}$& 705.009, 2283.42 & 1886.706, 6259.12 \\
$m_{\widetilde\chi^0_{3,4}}$& 3701.92,3702.45 & 7871.38, 7871.64 \\
$m_{\widetilde\chi^\pm_{1,2}}$& 705.182, 3705.43 & 1886.881, 7802.96 \\
$m_{\widetilde g}$& 4798 & 11984 \\
$m_{\widetilde u_{L,R}}$& 5210, 5239 & 13645, 13713 \\
$m_{\widetilde t_{1,2}}$& 3706, 4390 & 10385, 10636 \\
$m_{\widetilde d_{L,R}}$& 5211, 5280 & 13645, 13823 \\
$m_{\widetilde b_{1,2}}$& 4362, 4949 & 10574, 10949 \\ 
$m_{\widetilde\nu_1}$& 2555  & 7581 \\
$m_{\widetilde\nu_3}$& 2470 & 7180 \\
$m_{\widetilde e_{L,R}}$& 2559, 2519 & 7582, 7468 \\
$m_{\widetilde\tau_{1,2}}$& 2337, 2480 & 6575, 7177 \\
\hline\hline
\end{tabular}}
\caption{Two  benchmark points; point 1 and point 2 belong to red and orange solutions respectively. Masses are in GeV except for $M_{\rm GUT}$. The relic density is not imposed as a constraint.}
\label{table1}
\end{table}
\color{black}

\section{Conclusions}
\color{black}

We studied the SUSY GUTs with the minimal AMSB as a continuation of our program at the future 100 TeV proton-proton Collider such as $\rm FCC_{hh}$ and SppC as well as the proton-decay experiment such as Hyper-Kamiokande experiment. We showed that for the GUT scale $M_{GUT} \ge 0.84 \times 10^{16}$ GeV, the upper mass bounds on the gluino and first two generation squarks are 8 TeV and 6 TeV, respectively. Therefore, the gluino and first two generation squarks can be probed at the $\rm FCC_{hh}$ and SppC. In addition, we discuss the mass spectra for the colored sparticles, sleptons, pseudo-scalar Higgs and electroweakinos. We also shed light on the possible search strategies of gluino and winos at the future 100 TeV colliders. Moreover, for $\rm M_{\rm GUT} \le  0.84 \times 10^{16}$ GeV, the proton lifetime for dimension-six proton decay via heavy gauge boson exchange is less than or around $5 \times 10^{34}$ years, and thus we can probe it at the Hyper-Kamiokande experiment. To be concrete, we presented a couple of benchmark points representing the above two mentioned GUT conditions on $\rm M_{\rm GUT}$.

\color{black}
\clearpage

\begin{acknowledgments}
	
SR thanks TL and the Institute for Theoretical Physics, Chinese Academy of Sciences for the hospitality where part of this work was carried out.
TL is supported in part by the National Key Research and Development Program of China Grant No. 2020YFC2201504, by the Projects No. 11875062, No. 11947302, No. 12047503, and No. 12275333 supported by the National Natural Science Foundation of China, by the Key Research Program of the Chinese Academy of Sciences, Grant No. XDPB15, by the Scientific Instrument Developing Project of the Chinese Academy of Sciences, Grant No. YJKYYQ20190049, by the International Partnership Program of Chinese Academy of Sciences for Grand Challenges, Grant No. 112311KYSB20210012, and by the Henan Province Outstanding Foreign Scientist Studio Project, No.GZS2025008. This work was supported by the High Performance Computing Platform of Henan Normal University. The authors also acknowledge the
use of ChatGPT for assistance in improving the English
language and clarity of the manuscript. The authors remain
fully responsible for the scientific content, analysis, and
conclusions presented in this work

\end{acknowledgments}



\end{document}